\documentclass{iopart}
\usepackage{iopams}  
\usepackage{revsymb}  
\usepackage{graphicx,color}  
\usepackage{bm}  
\newcommand{\ket}[1]{|{#1}\rangle}
\newcommand{\bra}[1]{\langle{#1}|}
\newcommand{\bras}[2]{{}_{#2}\hspace*{-0.2mm}\langle{#1}|}
\newcommand{\ketbras}[3]{\ket{#1}_{#3}\hspace*{-0.2mm}\bra{#2}}
\newcommand{\bracket}[2]{\langle#1|#2\rangle}

\begin{document}
\title[Extraction of Entanglement by Resonant Transmission of Ancilla Qubit]{Extraction of an Entanglement by Repetition of the Resonant Transmission of an Ancilla Qubit}
\author{Kazuya Yuasa}
\address{Waseda Institute for Advanced Study, Waseda University, Tokyo 169-8050, Japan}
\ead{\mailto{yuasa@aoni.waseda.jp}}
\begin{abstract}
A scheme for the extraction of entanglement in two noninteracting qubits (spins) is proposed.
The idea is to make use of resonant transmission of ancilla qubit through the two fixed qubits, controlled by the entanglement in the scatterers.
Repetition of the resonant transmission extracts the singlet state in the target qubits from their arbitrary given state.
Neither the preparation nor the post-selection of the ancilla spin is required, in contrast to the previously proposed schemes.
\end{abstract}
\submitto{J. Phys. A: Math. Theor.}

\section{Introduction}
Entanglement is considered to be a key resource for quantum information technology \cite{ref:QuantumInfo}.
Its nonlocal character, which is truly quantum and beyond classical realm, is the driving force of the various attractive quantum information protocols.
Such a highly quantum state however is not easily found in laboratories, and its preparation itself is an important issue for the realization of the ideas of quantum information.

The entanglement would be most simply and naturally generated by a direct interaction between quantum systems, or between qubits in the context of quantum information \cite{ref:QdotQI}.
There would be however physical setups in which qubits are located separately beyond the range of the mutual interaction.
A possible way to make those qubits entangled is to send ``mediators'' to the qubits in order for them to communicate in an indirect manner through the mediators.
Such an idea has been explored in the literature 
\cite{ref:EntGeneMed,ref:qpfeSeparated,ref:qpfeScattering,ref:qpfesc-long,ref:EntReso,ref:qpfescCiccarelloPRL} and actually employed in several experiments \cite{ref:Haroche,ref:Photon}.

A typical scheme based on this idea is the following.
We consider two spin qubits $A$ and $B$ and take another spin as a mediator qubit $X$\@.
$A$ and $B$ are initially in a product state $\ket{\downarrow\downarrow}_{AB}$, and $X$ polarized in $\ket{\uparrow}_X$ is sent to make $A$ and $B$ entangled.
$X$ interacts first with $A$ and then with $B$\@.
Suppose that those interactions preserve the total spin angular momentum of the three qubits $XAB$\@.
Then, if $X$ is found to be flipped down to $\ket{\downarrow}_X$ after the interactions with $A$ and $B$, we are sure that the spin of either $A$ or $B$ is flipped up but we do not know which, and we end up with an entanglement $a\ket{\uparrow\downarrow}_{AB}+b\ket{\downarrow\uparrow}_{AB}$ between $A$ and $B$, with certain amplitudes $a$ and $b$\@.

In this scheme, the post-selection of the spin state of the mediator plays a crucial role.
There would be however various physical setups where such post-selection is technically hard.
For instance, if we consider an electron as a mediator, in the context of a solid-state system \cite{ref:qpfeScattering,ref:qpfesc-long,ref:EntReso,ref:qpfescCiccarelloPRL}, the post-selection of its spin state would be difficult by the current technology.

In addition, the above scheme requires the preparation of the target qubits in $\ket{\downarrow\downarrow}_{AB}$, before starting the protocol.
Preparation of a quantum state (not necessarily an entangled state) from an \textit{arbitrary} given state is also an important and nontrivial subject.
A generic scheme has been proposed to prepare a pure quantum state from an arbitrary (mixed, in general) state by repeated measurements on an ancilla system interacting with the target \cite{ref:qpfqpfe}. 
Applying this idea to the present issue, schemes for the extraction of entanglement between separated qubits, from an arbitrary given state \cite{ref:qpfeSeparated} or from a state belonging to a certain class of states \cite{ref:qpfescCiccarelloPRL}, by repeated measurements on the mediator have been explored.
Still, they require the preparation and the post-selection of the internal state of the mediator.

In this paper, we propose a scheme which is less demanding in these respects: neither the preparation nor the post-selection of the spin state of the mediator $X$ is required.
The initial state of the target qubits $A$ and $B$ can be arbitrary, from which the singlet state of $A$ and $B$ is extracted.
The idea is to make use of the resonant scattering of $X$ by $A$ and $B$, controlled by the entanglement in $A$ and $B$, which was discovered  in Ref.\ \cite{ref:EntReso}.
The resonant transmission of $X$ (which will be detailed below) acts as a filter, and the repetition of the resonant transmission extracts the singlet state of $A$ and $B$\@.

\section{Setup}
Suppose that two qubits $A$ and $B$ are fixed at $x=-d/2$ and $d/2$, respectively, along a 1D channel.
They do not interact directly with each other, while we wish to make them entangled.
In order to make them entangled, we send ancilla qubits $X$ to $A$ and $B$\@.
The initial state of $A$ and $B$ is arbitrary and is in general a mixed state $\rho_0$, from which we try to extract an entanglement.

\begin{figure}[b]
\begin{center}
\includegraphics[width=0.5\textwidth]{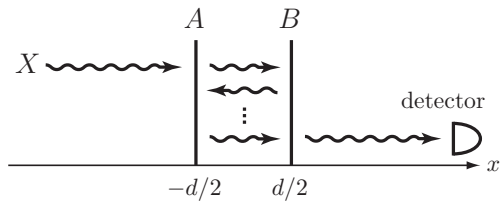}
\end{center}
\caption{
An ancilla qubit $X$ is sent to two fixed qubits $A$ and $B$ with a certain wave vector $k$, scattered by the delta-shaped potentials produced by $A$ and $B$, and is detected on the right side.
Neither the spin preparation of the incident $X$ nor the spin-resolved detection of the transmitted $X$ is required.}
\label{fig:setup}
\end{figure}
We consider the following Hamiltonian in 1D (sketched in Fig.\ \ref{fig:setup}):
\begin{equation}
H=\frac{p^2}{2m}
+g(\bm{\sigma}^{(X)}\cdot\bm{\sigma}^{(A)})\delta(x+d/2)
+g(\bm{\sigma}^{(X)}\cdot\bm{\sigma}^{(B)})\delta(x-d/2),
\label{eqn:Hamiltonian}
\end{equation}
where $x$ and $p$ are the position and the momentum of $X$, while $\bm{\sigma}^{(J)}\,(J=X,A,B)$ are the Pauli operators for the spins of $X$, $A$, and $B$\@.
The potentials produced by the fixed qubits $A$ and $B$ are represented by the delta-shaped potentials, and the spin-spin interaction between $X$ and each of $A$ and $B$ is of the Heisenberg type.
The spin states of each qubit are denoted by $\ket{\uparrow}$ and $\ket{\downarrow}$: the spin is oriented in the $z$ direction in the former, while in the opposite direction in the latter.

We send $X$ from the left with a certain incident wave vector $k\,(>0)$ and let it be scattered by $A$ and $B$\@.
The scattering matrix elements are given by (see the Appendix)
\begin{equation}
\bra{k'\zeta'}S\ket{k\zeta}
=e^{-ikd}[\delta(k'-k)\bra{\zeta'}T\ket{\zeta}
+\delta(k'+k)\bra{\zeta'}R\ket{\zeta}],
\label{eqn:SmatrixElements}
\end{equation}
where $\zeta$ represents the spin state of $XAB$, and $T$ and $R$ describe the changes induced in the spin state when $X$ is transmitted and reflected, respectively, given by
\numparts
\begin{eqnarray}
\fl
T
=e^{ikd}\,\Bigl[
\alpha(1-4 i\Omega)P_-
+(\alpha Q_\frac{1}{2}+\beta Q_\frac{3}{2})P_+
\nonumber\\
\fl\qquad\qquad\qquad\quad
{}-\alpha\Omega^2(1-e^{2 ikd})
(
P_-
-3Q_\frac{1}{2}P_+
-K_-+K_+
)
\Bigr],
\label{eqn:TRa}
\end{eqnarray}
\begin{eqnarray}
\fl
R
=T e^{- ikd}-1
-i\Omega(1-e^{2 ikd}) 
\,\biggl\{
6\alpha\Omega^2(1-e^{2 ikd})P_-
+(2\alpha Q_\frac{1}{2}
-\beta Q_\frac{3}{2})P_+
\nonumber\\
\fl\qquad\qquad\qquad\quad
{}+\frac{1}{2}\alpha
(K_++K_-)[
1+3\Omega^2(1-e^{2 ikd})
-4 i\Omega P_-
]
\biggr\},
\label{eqn:TRb}
\end{eqnarray}
\endnumparts
with
\numparts
\begin{eqnarray}
\alpha =\frac{1}{\displaystyle
(1-4  i\Omega) +2 \Omega ^2(1-6 i\Omega) (1- e^{2ikd})+9\Omega ^4 (1- e^{2ikd})^2 
},\\
\beta =\frac{1}{(1+2  i\Omega) -\Omega ^2(1- e^{2ikd}) },\qquad
\Omega=\frac{mg}{\hbar^2k}.
\label{eqn:AlphaBetaOmega}
\end{eqnarray}
\endnumparts
Here,
\begin{equation}
P_-
=\frac{1-\bm{\sigma}^{(A)}\cdot\bm{\sigma}^{(B)}}{4}
,\qquad
P_+
=\frac{3+\bm{\sigma}^{(A)}\cdot\bm{\sigma}^{(B)}}{4}
\label{eqn:Ppm}
\end{equation}
are the projection operators on the singlet and triplet sectors of $A$ and $B$, respectively, while
\numparts
\begin{eqnarray}
Q_\frac{3}{2}=\frac{2}{3}P_++\frac{1}{6}\bm{\sigma}^{(X)}\cdot(\bm{\sigma}^{(A)}+\bm{\sigma}^{(B)}),
\label{eqn:Qa}
\\
Q_\frac{1}{2}=P_-+\frac{1}{3}P_+-\frac{1}{6}\bm{\sigma}^{(X)}\cdot(\bm{\sigma}^{(A)}+\bm{\sigma}^{(B)})
\label{eqn:Qb}
\end{eqnarray}
\endnumparts
are those on the spin-$\frac{3}{2}$ and spin-$\frac{1}{2}$ sectors of $XAB$, respectively \cite{ref:qpfesc-long}. 
Note that these projection operators are all commutative with each other and $Q_\frac{3}{2}P_-=P_-Q_\frac{3}{2}=0$.
The other operators 
\begin{equation}
K_\pm=\frac{1}{2}\bm{\sigma}^{(X)}\cdot[(\bm{\sigma}^{(A)}-\bm{\sigma}^{(B)})
\pm i(\bm{\sigma}^{(A)}\times\bm{\sigma}^{(B)})]
\end{equation}
describe the transitions between the singlet and triplet sectors of $A$ and $B$, with the only nonzero elements $P_\pm K_\pm P_\mp\neq0$.
The unitarity of the scattering matrix is expressed as
\begin{equation}
T^\dag T+R^\dag R=\openone_{XAB}.
\label{eqn:Unitarity}
\end{equation}

\begin{figure}
\begin{center}
\includegraphics[width=0.5\textwidth]{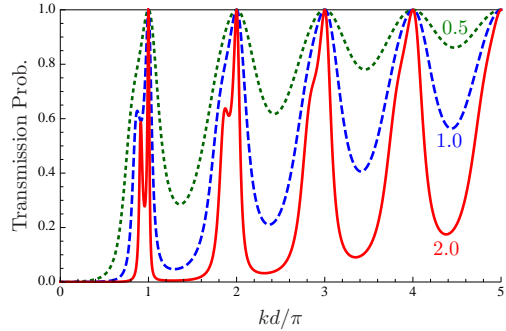}
\end{center}
\caption{The transmission probability of $X$, incident with its spin completely mixed $\openone_X/2$ toward the scatterers $A$ and $B$ in the singlet state $\ket{\Psi^-}_{AB}$, as a function of the incident wave vector $k$, for different coupling constants $mgd/\hbar^2\pi=0.5\,\mathrm{(dotted)},\,1.0\,\mathrm{(dashed)},\,2.0\,\mathrm{(solid)}$.}
\label{fig:Reso}
\end{figure}
It is clarified in Ref.\ \cite{ref:EntReso} that this system (with the same coupling constant $g$ for $A$ and $B$) exhibits interesting resonant transmission depending on the state of $A$ and $B$\@.
For instance, perfect transmission is realized at certain incident momenta, when $A$ and $B$ are in an entangled state $\ket{\Psi^-}_{AB}=(\ket{\uparrow\downarrow}_{AB}-\ket{\downarrow\uparrow}_{AB})/\sqrt{2}$ and $X$ is sent in an arbitrary spin state.
See Fig.\ \ref{fig:Reso}.
In fact, we have
\numparts
\begin{eqnarray}
T
=(-1)^n\left[
P_-
+\left(
\frac{1}{1-4 i\Omega}Q_\frac{1}{2}
+\frac{1}{1+2  i\Omega}Q_\frac{3}{2}
\right)
P_+
\right],
\label{eqn:TRreso-a}
\\
R=\left(
\frac{4 i\Omega}{1-4  i\Omega}
Q_\frac{1}{2}
-\frac{2 i\Omega}{1+2  i\Omega}
Q_\frac{3}{2}
\right)
P_+
\label{eqn:TRreso-b}
\end{eqnarray}
\endnumparts
at $k=n\pi/d\,(n=1,2,\ldots)$, which show that $X$ is perfectly transmitted without spin flip, provided that $A$ and $B$ are in the single state $\ket{\Psi^-}_{AB}$.

By making use of this resonance, a scheme for extracting the singlet state $\ket{\Psi^-}_{AB}$ in $A$ and $B$ is proposed in Ref.\ \cite{ref:qpfescCiccarelloPRL}.
$X$ prepared in $\ket{\uparrow}_{X}$ is injected with a resonant wave vector $k=n\pi/d\,(n=1,2,\ldots)$, and the same state $\ket{\uparrow}_{X}$ of $X$ is post-selected on the left (reflected) or on the right (transmitted) after the scattering by $A$ and $B$\@. 
Repetition of this process extracts the singlet state $\ket{\Psi^-}_{AB}$ from a state belonging to a certain class of states of $A$ and $B$: the state of $A$ and $B$ need not be prepared in a specific state before starting the protocol.
In this paper, we propose a different scheme that works for an \textit{arbitrary} initial state of $A$ and $B$\@. Furthermore, it requires neither the preparation nor the post-selection of the spin state of $X$, in contrast to the scheme proposed in Ref.\ \cite{ref:qpfescCiccarelloPRL}.
Preparing and post-selecting the spin state, e.g.\ of an electron in solid, are technically hard in practice.
The scheme we are going to discuss in the following has an advantage over the previous schemes in this respect.

\section{Protocol}
Our procedure is the following:
\begin{enumerate}
\setcounter{enumi}{-1}
\item The initial state of $A$ and $B$ is \textit{arbitrary} and is a mixed state $\rho_0$ in general.
\item We send $X$ with its spin \textit{arbitrary} (which can be ``random'') from the left to $A$ and $B$, with a resonant wave vector $k=n\pi/d\,(n=1,2,\ldots)$, and detect it on the right after the scattering, \textit{without resolving the spin state of X}.
\item We repeat this process many times, and if $X$ is repeatedly confirmed to be transmitted to the right (\textit{irrespectively of its spin state}), the singlet state $\ket{\Psi^-}_{AB}$ (an entanglement) is extracted in $A$ and $B$ from $\rho_0$.
\end{enumerate}

Once $X$ is confirmed to be reflected in a trial, we stop the procedure: we fail to extract the singlet state.
The singlet state is obtained only when all the $X$s sent are confirmed to be transmitted.
This is a probabilistic scheme for the extraction of the singlet state.

In the scheme proposed in Ref.\ \cite{ref:qpfescCiccarelloPRL}, the incident spin of $X$ is prepared in $\ket{\uparrow}_X$ and the same state $\ket{\uparrow}_X$ is post-selected after the scattering (irrespectively of whether it is transmitted or reflected).
In the present scheme, on the other hand, neither the preparation nor the post-selection of the incident spin is required.
Instead, we post-select the transmission events.
The singlet spin state of the pair of qubits is extracted by the repetition of the post-selection for the spatial degree of freedom of the mediator.

If the incident spin of $X$ is not polarized and is completely random, it would be represented by the completely mixed state $\openone_{X}/2$ on average.
Then, the confirmation of the transmission of $X$ induces the following change in the spin state of $A$ and $B$, 
\begin{equation}
\rho_0
\to\mathcal{T}\rho_0
=\Tr_{X}\{
T
(\openone_{X}/2\otimes\rho_0)T^\dag
\},
\label{eqn:MapT}
\end{equation}
up to normalization.
Notice that, since we do not resolve the spin state of $X$ after the transmission, we take trace over possible spin states of $X$\@.
$N$ repetitions of this process drive $A$ and $B$ into
\begin{equation}
\rho(N)
=\mathcal{T}^N\rho_0/P(N),\qquad
P(N)=\Tr_{AB}\{\mathcal{T}^N\rho_0\}.
\label{eqn:ExtState}
\end{equation}
Here the extracted state $\rho(N)$ is normalized to unity.
This state $\rho(N)$ is obtained only when the transmission of $X$ is confirmed successively $N$ times, and the normalization factor $P(N)$ gives the probability for this to occur. 
We are going to prove that $A$ and $B$ are driven into the singlet state
\begin{equation}
\rho(N)
\to\ketbras{\Psi^-}{\Psi^-}{{AB}}
\quad\mathrm{as}\quad
N\to\infty
\label{eqn:Purification}
\end{equation}
after the repeated confirmations of the transmissions, and the probability approaches 
\begin{equation}
P(N)
\to
\bras{\Psi^-}{{AB}}\rho_0\ket{\Psi^-}_{AB}
\quad\mathrm{as}\quad
N\to\infty
.
\label{eqn:Prob}
\end{equation}
The singlet state $\ket{\Psi^-}_{AB}$ is extracted from an arbitrary given state $\rho_0$ with a \textit{nonvanishing} probability $\bras{\Psi^-}{{AB}}\rho_0\ket{\Psi^-}_{AB}$, as long as the initial state $\rho_0$ has a singlet component.

\section{Proof}
By looking at the explicit expression of $T$ at resonance $k=n\pi/d\,(n=1,2,\ldots)$, given in (\ref{eqn:TRreso-a}), it is easy to check that the singlet state $\ketbras{\Psi^-}{\Psi^-}{AB}$ is a fixed point of the map $\mathcal{T}$ and the projection onto the singlet state
\begin{equation}
\mathcal{P}_-\rho
=P_-\rho P_-
\end{equation}
is the eigenprojection of $\mathcal{T}$ belonging to the eigenvalue $1$,
\begin{equation}
\mathcal{T}\mathcal{P}_-
=\mathcal{P}_-\mathcal{T}
=\mathcal{P}_-.
\end{equation}
The convergence (\ref{eqn:Purification}) is proved by showing that the singlet state is the only eigenvector belonging to the eigenvalue $1$ and all the other eigenvalues are strictly smaller than $1$.
The proof proceeds similarly to a proof found in \cite{ref:GeneLyapunov}.

\begin{figure}[t]
\begin{center}
\includegraphics[width=0.5\textwidth]{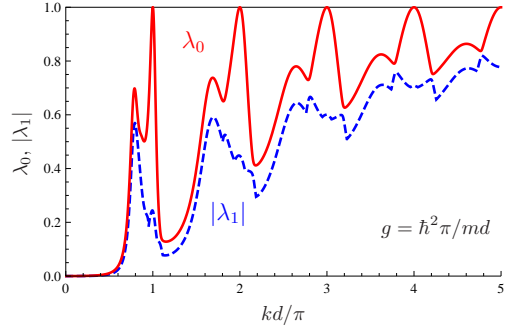}
\end{center}
\caption{The largest and the second largest eigenvalues (in magnitude), $\lambda_0$ and $\lambda_1$, respectively, of the map $\mathcal{T}$, as functions of the incident wave vector $k$ of $X$\@. At the resonance points $kd/\pi=1,2,\ldots$, the largest eigenvalue reaches $\lambda_0=1$, while $|\lambda_1|$ is certainly smaller than $1$. The coupling constant is fixed at $g=\hbar^2\pi/md$.}
\label{fig:Eigenvalues}
\end{figure}

Suppose that the map $\mathcal{T}$ admits an eigenvalue of unit magnitude (\textit{peripheral eigenvalue}),
\begin{equation}
\mathcal{T}\nu=e^{i\varphi}\nu,
\label{eqn:UnitEigen}
\end{equation}
where $\nu$ is the eigenvector belonging to the eigenvalue $e^{i\varphi}$ with a real number $\varphi$.
The polar decomposition of $\nu$ is always possible,
\begin{equation}
\nu=U\sqrt{\nu^\dag\nu}
=\gamma U\sigma,
\label{eqn:polar}
\end{equation}
where $\sigma=\sqrt{\nu^\dag\nu}/\gamma$ is a normalized density operator with the normalization constant $\gamma=\Tr_{AB}\sqrt{\nu^\dag\nu}$, and $U$ is a unitary operator.
By inserting (\ref{eqn:polar}) into (\ref{eqn:UnitEigen}), we have
\begin{equation}
\mathcal{T}(U\sigma)=e^{i\varphi}U\sigma,
\end{equation}
which yields
\begin{equation}
e^{i\varphi}
=\Tr_{AB}\{U^\dag\mathcal{T}(U\sigma)\}.
\end{equation}
Then, recalling the definition of $\mathcal{T}$ in (\ref{eqn:MapT}), 
\begin{eqnarray}
1&=|{\Tr_{AB}\{U^\dag\mathcal{T}(U\sigma)\}}|
\nonumber
\\
&=|{\Tr\{U^\dag T(U\sigma/2)T^\dag\}}|.
\nonumber
\\
&=|{\Tr\{(\sigma/2)T^\dag(U^\dag TU)\}}|.
\nonumber
\\
&\le
\sqrt{{\Tr\{(\sigma/2)T^\dag T\}}}
\sqrt{{\Tr\{(\sigma/2)U^\dag T^\dag TU\}}}
\nonumber
\\
&=
\sqrt{\Tr_{AB}\{\mathcal{T}\sigma\vphantom{U^\dag}\}}
\sqrt{\Tr_{AB}\{\mathcal{T}(U\sigma U^\dag)\}}
\le1,
\label{eqn:IEQ1}
\end{eqnarray}
where the Cauchy-Schwarz inequality is used by noting $\openone_X/2\otimes\sigma$ is a state of $XAB$ normalized to unity.
The inequality (\ref{eqn:IEQ1}) implies
\begin{equation}
\Tr_{AB}\{\mathcal{T}\sigma\}
=\Tr_{AB}\{\mathcal{T}(U\sigma U^\dag)\}=1.
\label{eqn:Cond1}
\end{equation}
Let us look at the first condition.
By noting the unitarity (\ref{eqn:Unitarity}), it is written as
\begin{equation}
\Tr_{AB}\{\mathcal{T}\sigma\}
=1-\Tr\{(\sigma/2)R^\dag R\}=1,
\end{equation}
and hence, is reduced to
\begin{equation}
\Tr\{(\sigma/2)R^\dag R\}=0.
\end{equation}
By inserting the explicit expression of $R$ at resonance $k=n\pi/d\,(n=1,2,\ldots)$, given in (\ref{eqn:TRreso-b}), it reads
\begin{equation}
\Tr\{(\sigma/2)R^\dag R\}
=\frac{8\Omega^2(1+8\Omega^2)}{(1+4\Omega^2)(1+16\Omega^2)}
\Tr_{AB}\{P_+\sigma\}
=0.
\end{equation}
Since this coefficient is nonvanishing, this implies
\begin{equation}
\Tr_{AB}\{P_+\sigma\}=0,
\end{equation} 
and further,
\begin{equation}
\sigma=\ketbras{\Psi^-}{\Psi^-}{AB}.
\label{eqn:Sigma1}
\end{equation}
Similarly, the second condition in (\ref{eqn:Cond1}) implies
\begin{equation}
U\sigma U^\dag=\ketbras{\Psi^-}{\Psi^-}{AB}.
\label{eqn:Sigma2}
\end{equation}
Combination of the conditions (\ref{eqn:Sigma1}) and (\ref{eqn:Sigma2}) yields
\begin{equation}
U\ketbras{\Psi^-}{\Psi^-}{AB}U^\dag=\ketbras{\Psi^-}{\Psi^-}{AB},
\end{equation}
and hence,
\begin{equation}
U\ket{\Psi^-}_{AB}=e^{i\xi}\ket{\Psi^-}_{AB}
\label{eqn:U}
\end{equation}
with a real number $\xi$.
Inserting (\ref{eqn:Sigma1}) and (\ref{eqn:U}) to (\ref{eqn:polar}), we obtain 
\begin{equation}
\nu=U\ketbras{\Psi^-}{\Psi^-}{AB}=e^{i\xi}\ketbras{\Psi^-}{\Psi^-}{AB},
\end{equation}
which yields $\varphi=0$ for the eigenvalue in (\ref{eqn:UnitEigen}).
Since the phase $e^{i\xi}$ in the eigenvector is irrelevant to the eigenvalue problem, this proves that the singlet state $\ketbras{\Psi^-}{\Psi^-}{AB}$ is the only eigenvector of the map $\mathcal{T}$ belonging to a peripheral eigenvalue, which is actually $1$, and all the other eigenvalues are strictly smaller than $1$.
See Fig.\ \ref{fig:Eigenvalues}, where the magnitudes of the two largest eigenvalues of the map $\mathcal{T}$ are shown as functions of the incident wave vector $k$.

Then, split the map $\mathcal{T}$ into two parts as
\begin{equation}
\mathcal{T}=\mathcal{P}_-+\mathcal{T}'.
\end{equation}
Since the eigenprojections belonging to different eigenvalues are orthogonal to each other,
\begin{equation}
\mathcal{P}_-\mathcal{T}'
=\mathcal{T}'\mathcal{P}_-=0.
\end{equation}
In addition, $\mathcal{T}'^N\to0$ as $N\to\infty$, since the spectral radius of $\mathcal{T}'$ is strictly smaller than $1$.
Therefore, the map $\mathcal{T}^N$ converges to
\begin{equation}
\mathcal{T}^N=\mathcal{P}_-+\mathcal{T}'^N
\to
\mathcal{P}_-
\quad\mathrm{as}\quad
N\to\infty
,
\end{equation}
which proves (\ref{eqn:Purification}) and (\ref{eqn:Prob}).

\section{Fidelity}
\label{sec:Fidelity}
It is possible to compute the evolution of the fidelity of the extracted state $\rho(N)$ with respect to the target singlet state $\ket{\Psi^-}_{AB}$,
\begin{equation}
F(N)=\Tr_{AB}\{P_-\rho(N)\},
\end{equation}
and the probability $P(N)$.
In fact, by inserting the explicit expression of $T$ given in (\ref{eqn:TRa}) into the definition of the map $\mathcal{T}$ in (\ref{eqn:MapT}), one realizes that the transitions between the singlet and triplet sectors provoked by $\mathcal{T}$ are described by 
\begin{equation}
\left(
\begin{array}{c}
\medskip
\displaystyle
(\mathcal{T}\rho)_-\\
\displaystyle
(\mathcal{T}\rho)_+
\end{array}
\right)
=\left(
\begin{array}{cc}
\medskip
\displaystyle
\mathcal{T}_{--}&
\mathcal{T}_{-+}\\
\mathcal{T}_{+-}&
\mathcal{T}_{++}
\end{array}
\right)
\left(
\begin{array}{cc}
\medskip
\displaystyle
\rho_-\\
\displaystyle
\rho_+
\end{array}
\right),
\label{eqn:Transition}
\end{equation}
where
\begin{equation}
\rho_\pm=\Tr_{AB}\{P_\pm\rho\}
\end{equation}
and
\numparts
\begin{eqnarray}
\mathcal{T}_{--}
={}&|\alpha|^2
|
1-4i\Omega-\Omega^2(1-e^{2ikd})
|^2,
\\
\mathcal{T}_{-+}
={}&\frac{1}{3}\mathcal{T}_{+-}
=4|\alpha|^2\Omega^4
|
1-e^{2ikd}
|^2,
\\
\mathcal{T}_{++}
={}&\frac{1}{9}|
(\alpha+2\beta)
+3\alpha\Omega^2(1-e^{2ikd})
|^2
\nonumber\\
&{}+\frac{2}{9}|
(\alpha-\beta)
+3\alpha\Omega^2(1-e^{2ikd})
|^2.
\end{eqnarray}
\endnumparts
For $\mathcal{T}^N$, one has
\begin{equation}
\left(
\begin{array}{c}
\medskip
\displaystyle
(\mathcal{T}^N\rho_0)_-\\
\displaystyle
(\mathcal{T}^N\rho_0)_+
\end{array}
\right)
=\left(
\begin{array}{cc}
\medskip
\displaystyle
\mathcal{T}_{--}&
\mathcal{T}_{-+}\\
\mathcal{T}_{+-}&
\mathcal{T}_{++}
\end{array}
\right)^N
\left(
\begin{array}{c}
\medskip
\displaystyle
(\rho_0)_-\\
\displaystyle
(\rho_0)_+
\end{array}
\right).
\end{equation}
At the resonance points $k=n\pi/d\,(n=1,2,\ldots)$, these matrix elements are reduced to 
\begin{eqnarray}
\mathcal{T}_{--}=1,\quad
\mathcal{T}_{-+}=\mathcal{T}_{+-}=0,\quad
\mathcal{T}_{++}=\frac{1+12\Omega^2}{(1+4\Omega^2)(1+16\Omega^2)},
\end{eqnarray}
and concise expressions of the fidelity and the probability are readily available,
\begin{equation}
F(N)
=(\rho_0)_-/P(N),
\end{equation}
\begin{equation}
P(N)
=(\rho_0)_-+\left(\frac{1+12\Omega^2}{(1+4\Omega^2)(1+16\Omega^2)}\right)^N(\rho_0)_+,
\end{equation}
which clearly show that the fidelity approaches $F(N)\to1$ as $N\to\infty$ and the singlet state $\ket{\Psi^-}_{AB}$ is extracted with probability $P(N)\to(\rho_0)_-$, proving (\ref{eqn:Purification}) and (\ref{eqn:Prob}).

\begin{figure}
\begin{center}
\begin{tabular}{l@{\ \ }l}
\footnotesize(a)&\footnotesize(b)\\[-3.5truemm]
\,%
\includegraphics[width=0.47\textwidth]{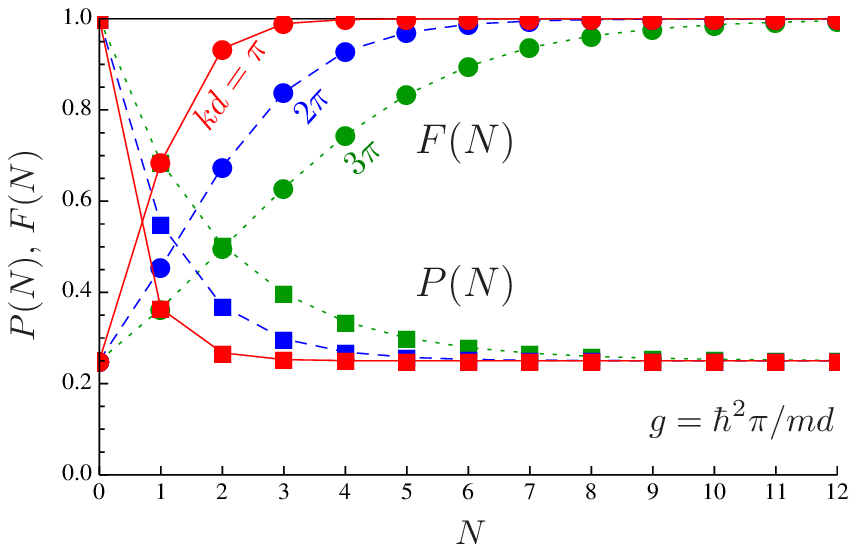}
&\,%
\includegraphics[width=0.47\textwidth]{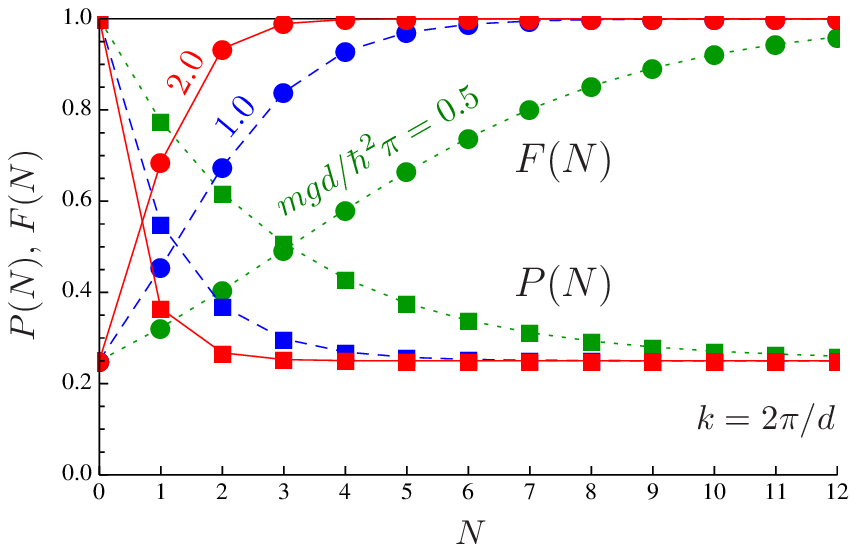}
\end{tabular}
\end{center}
\caption{The fidelity $F(N)$ of the extracted state $\rho(N)$ with respect to the singlet state $\ket{\Psi^-}_{AB}$ and the probability $P(N)$ of successive resonant transmissions, as functions of the number of repetitions $N$, (a) for different resonant wave vectors $kd/\pi=1\,\mathrm{(solid)},\,2\,\mathrm{(dashed)},\,3\,\mathrm{(dotted)}$ with $g=\hbar^2\pi/md$, and (b) for different coupling constants $mgd/\hbar^2\pi=0.5\,\mathrm{(dotted)},\,1.0\,\mathrm{(dashed)},\,2.0\,\mathrm{(solid)}$ with $k=2\pi/d$. The initial state of $A$ and $B$ is the completely mixed state $\rho_0=\openone_{AB}/4$ for both panels.}
\label{fig:FP}
\end{figure}
See Fig.\ \ref{fig:FP}, where the extraction of the singlet state $\ket{\Psi^-}_{AB}$ is demonstrated from the completely mixed state $\rho_0=\openone_{AB}/4$, with different resonant wave vectors $k$ and different coupling constants $g$.
The fidelity $F(N)$ approaches $1$ after several repetitions of the resonant transmissions, and the probability decays monotonically to $0.25$, which is the overlap between the initial state $\rho_0=\openone_{AB}/4$ and the target $\ket{\Psi^-}_{AB}$.
The speed of the extraction is ruled by the ratio $|\lambda_1/\lambda_0|$ between the largest and the second largest eigenvalues shown in Fig.\ \ref{fig:Eigenvalues} \cite{ref:qpfqpfe}: the extraction is faster with a smaller resonant wave vector $k$ and a larger coupling constant $g$.

\section{Comparison}
Let us compare the present scheme with the previously proposed scheme in Ref.\ \cite{ref:qpfescCiccarelloPRL}.
In the latter scheme, the incident spin of $X$ is prepared in $\ket{\uparrow}_X$ and the same spin state $\ket{\uparrow}_X$ is post-selected after the scattering by $A$ and $B$, irrespectively of whether $X$ is transmitted or reflected.
In the present scheme, on the other hand, the incident spin state of $X$ can be arbitrary and its spin state after the scattering is not checked.
Let us see how this difference affects the efficiency of the protocols.

\begin{figure}[b]
\begin{center}
\includegraphics[width=0.5\textwidth]{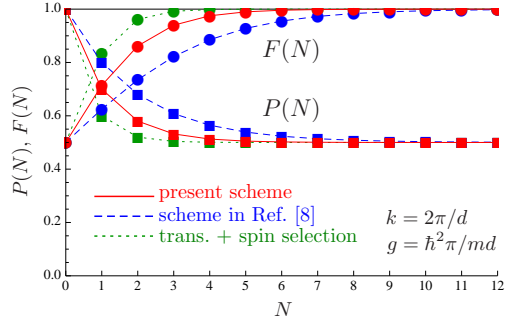}
\end{center}
\caption{Comparison of three different schemes, i.e.\ the present scheme (solid line: with post-selection of transmission but without preparation or post-selection of spin), the scheme proposed in Ref.\ \cite{ref:qpfescCiccarelloPRL} (dashed line: without post-selection of transmission or reflection but with preparation and post-selection of spin), and another (dotted line: with post-selection of transmission as well as with preparation and post-selection of spin).
The initial state of $A$ and $B$ is a product state $\rho_0=\ketbras{\uparrow\downarrow}{\uparrow\downarrow}{AB}$, and the parameters are chosen to be $k=2\pi/d$, $g=\hbar^2\pi/md$.}
\label{fig:FP_Ciccarello}
\end{figure}
The performances of the two schemes are compared in Fig.\ \ref{fig:FP_Ciccarello}.
Recall first that the scheme proposed in Ref.\ \cite{ref:qpfescCiccarelloPRL} does not work if the initial state of $A$ and $B$ contains $\ket{\uparrow\uparrow}_{AB}$ component.
The extraction of the singlet state $\ket{\Psi^-}_{AB}$ is therefore demonstrated in Fig.\ \ref{fig:FP_Ciccarello} for the initial state given by $\rho_0=\ketbras{\uparrow\downarrow}{\uparrow\downarrow}{AB}$.\footnote{Note that the scheme proposed in Ref.\ \cite{ref:qpfescCiccarelloPRL} is intended to extract the singlet state of two spin-$s$ particles with $s$ in general greater than $1/2$.}
As Fig.\ \ref{fig:FP_Ciccarello} shows, the convergence to the singlet state $\ket{\Psi^-}_{AB}$ is faster in the present scheme than in the scheme proposed in Ref.\ \cite{ref:qpfescCiccarelloPRL}.
This is because reflection events are collected in the latter scheme.
The singlet state $\ket{\Psi^-}_{AB}$ in $A$ and $B$ never provokes the reflection of $X$, and therefore, the reflection of $X$, if it happens, is a signature of the presence of the triplet components in $A$ and $B$, which are to be cut away to extract the singlet state $\ket{\Psi^-}_{AB}$.
By keeping the reflection events, those unnecessary triplet components are retained in $A$ and $B$.
The present scheme, on the other hand, collects only transmission events: the reflection events are projected out, and so are some portions of the triplet components.
That is why the distillation of the singlet state $\ket{\Psi^-}_{AB}$ is faster in the present scheme.
The decay of the success probability is also faster, but the asymptotic values are the same for both schemes.

If we introduce the preparation and the post-selection of the spin state of $X$ in the present scheme, say preparing the incident spin in $\ket{\uparrow}_X$ and post-selecting $\ket{\uparrow}_X$ of the \textit{transmitted} $X$, the convergence to the singlet state $\ket{\Psi^-}_{AB}$ becomes quicker.
See Fig.\ \ref{fig:FP_Ciccarello} again.
As is clear from (\ref{eqn:TRreso-a}), $X$ is certainly transmitted without spin flip when $A$ and $B$ are in the singlet state $\ket{\Psi^-}_{AB}$.
If the spin of $X$ is found to be flipped after the transmission, this captures the presence of the triplet components in $A$ and $B$.
Therefore, by post-selecting $X$ transmitted without spin flip, some amount of triplet components are projected out, and the singlet state $\ket{\Psi^-}_{AB}$ is extracted more efficiently. 
In this way, the spin-resolved detection can enhance the efficiency of the protocol, although such detection would be technologically more demanding.

\section{Robustness}
The present scheme assumes that the incident wave of $X$ is a plane wave, monochromatized to a resonant momentum.
If this condition is violated, the scheme may fail to extract the singlet state.
The incident particle $X$ would be represented by a wave packet of finite width, or the incident momentum may fluctuate around the resonant momentum.
Let us finally discuss how such imperfections affect the scheme \cite{ref:qpfescCiccarelloIJQI}.

\begin{figure}[b]
\begin{center}
\begin{tabular}{l@{\ }l}
\footnotesize(a)&\footnotesize(b)\\[-3.5truemm]
\includegraphics[width=0.48\textwidth]{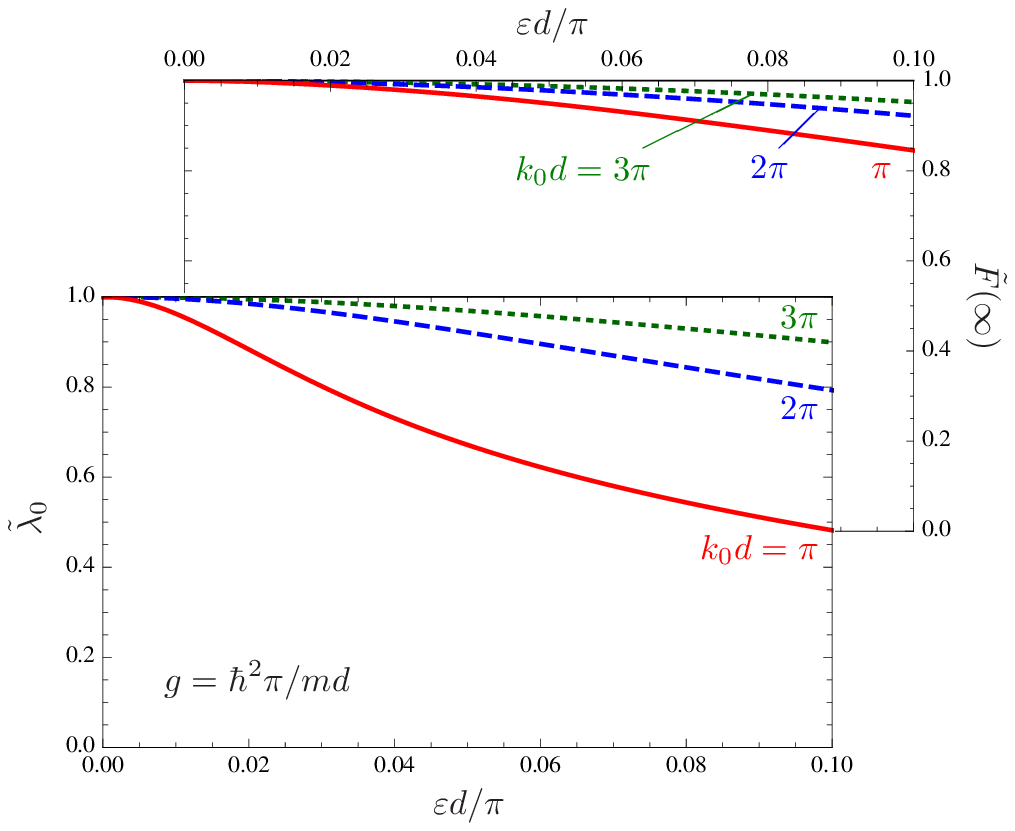}&
\includegraphics[width=0.48\textwidth]{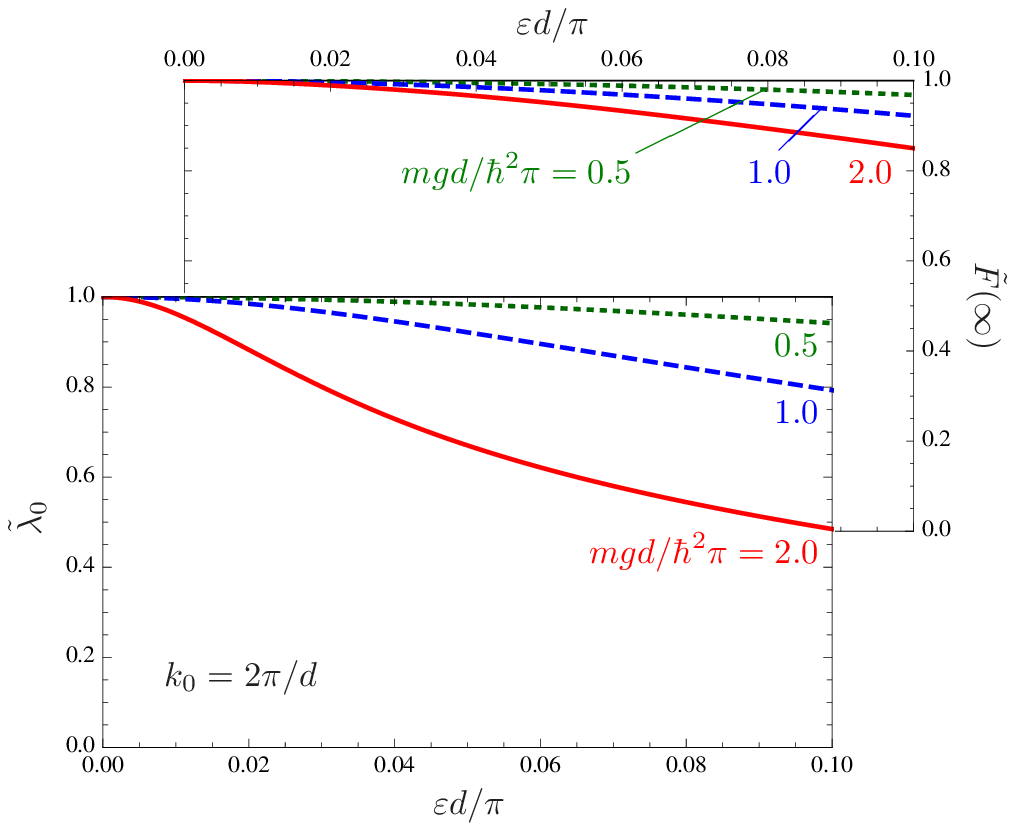}
\end{tabular}
\end{center}
\caption{The fidelity $\tilde{F}(\infty)$ of the extracted state $\tilde{\rho}(\infty)$ with respect to the singlet state $\ket{\Psi^-}_{AB}$ and the largest eigenvalue $\tilde{\lambda}_0$ of the map $\tilde{\mathcal{T}}$, as functions of the width $\varepsilon$ of the Gaussian distribution of the incident wave vector, (a) for different central wave vectors $k_0d/\pi=1\,\mathrm{(solid)},\,2\,\mathrm{(dashed)},\,3\,\mathrm{(dotted)}$ with $g=\hbar^2\pi/md$, and (b) for different coupling constants $mgd/\hbar^2\pi=0.5\,\mathrm{(dotted)},\,1.0\,\mathrm{(dashed)},\,2.0\,\mathrm{(solid)}$ with $k_0=2\pi/d$.}
\label{fig:Fluctuation}
\end{figure}
Suppose that the incident particle $X$ is not perfectly monochromatized but is represented by a wave packet $\psi(k)$ in momentum space.
In this case, the map $\mathcal{T}$ defined in (\ref{eqn:MapT}) with a fixed wave vector $k$ is replaced by
\begin{equation}
\tilde{\mathcal{T}}
=\int_0^\infty dk\,|\psi(k)|^2\mathcal{T}.
\end{equation}
The trace over the momentum degree of freedom is taken, since the detector does not resolve the momentum of the transmitted particle.
Notice that the effect of incoherent fluctuation of the incident momentum is also described by the same formula but with the packet $|\psi(k)|^2$ replaced by a probability distribution $f(k)$ characterizing the fluctuation.

\begin{figure}
\begin{center}
\begin{tabular}{l@{\ \ }l}
\footnotesize(a)&\footnotesize(b)\\[-3.5truemm]
\,%
\includegraphics[width=0.47\textwidth]{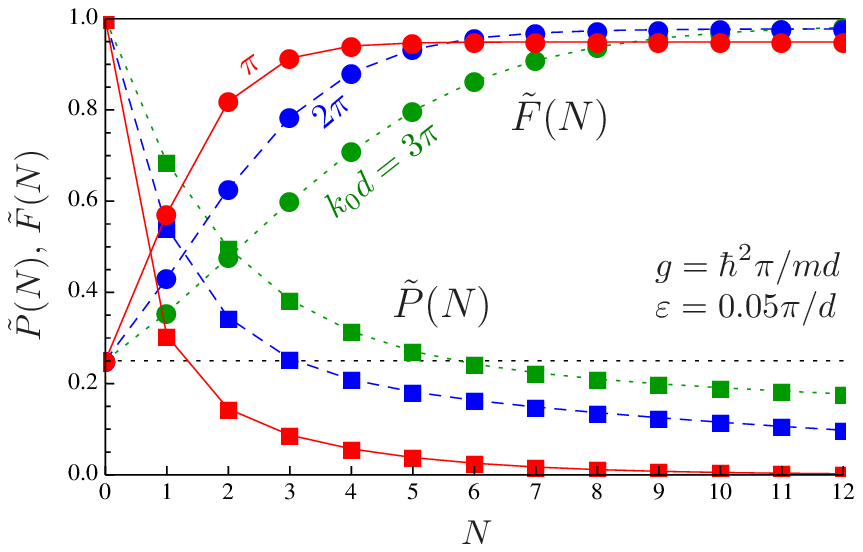}
&\,%
\includegraphics[width=0.47\textwidth]{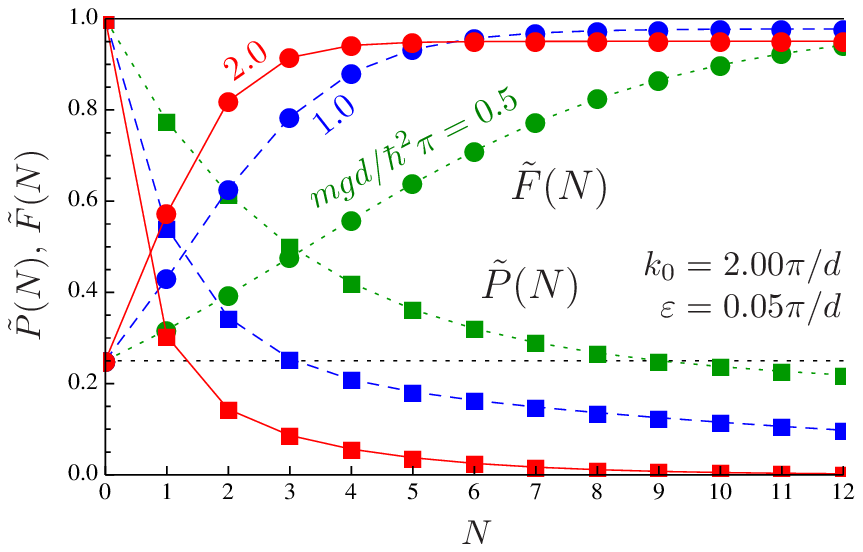}
\end{tabular}
\end{center}
\caption{The fidelity $\tilde{F}(N)$ of the extracted state $\tilde{\rho}(N)$ with respect to the singlet state $\ket{\Psi^-}_{AB}$ and the probability $\tilde{P}(N)$ of successive transmissions as functions of the number of repetitions $N$, with the width of the Gaussian distribution of the incident wave vectors, $\varepsilon=0.05\pi/d$, (a) for different central wave vectors $k_0d/\pi=1\,\mathrm{(solid)},\,2\,\mathrm{(dashed)},\,3\,\mathrm{(dotted)}$ with $g=\hbar^2\pi/md$, and (b) for different coupling constants $mgd/\hbar^2\pi=0.5\,\mathrm{(dotted)},\,1.0\,\mathrm{(dashed)},\,2.0\,\mathrm{(solid)}$ with $k_0=2\pi/d$. The initial state of $A$ and $B$ is the completely mixed state $\rho_0=\openone_{AB}/4$ for both panels.
The horizontal dotted line indicates the asymptotic value of the probability $0.25$ for the ideal case with $\varepsilon=0$.}
\label{fig:FPF}
\end{figure}
Now, the state of $A$ and $B$ after $N$ applications of the map $\tilde{\mathcal{T}}$ reads
\begin{equation}
\tilde{\rho}(N)
=\tilde{\mathcal{T}}^N\rho_0/\tilde{P}(N),\qquad
\tilde{P}(N)=\Tr_{AB}\{\tilde{\mathcal{T}}^N\rho_0\}.
\end{equation}
Let us look at the fidelity of the extracted state $\tilde{\rho}(N)$ with respect to the target singlet state,
\begin{equation}
\tilde{F}(N)
=\Tr_{AB}\{P_-\tilde{\rho}(N)\}.
\end{equation}
The asymptotic value $\tilde{F}(\infty)$ is readily evaluated from the eigenvector belonging to the larger eigenvalue $\tilde{\lambda}_0$ of the $2\times2$ matrix in (\ref{eqn:Transition}) with each matrix element integrated over $k$ with the relevant weight function $|\psi(k)|^2$ or $f(k)$.
We assume a Gaussian distribution $|\psi(k)|^2$ or $f(k)=e^{-(k-k_0)^2/2\varepsilon^2}/\sqrt{2\pi\varepsilon^2}$, centered at a resonant wave vector $k_0=n\pi/d\,(n=1,2,\ldots)$ with a band width $\varepsilon$, and the fidelity $\tilde{F}(\infty)$ is plotted as a function of $\varepsilon$ in Fig.\ \ref{fig:Fluctuation}.
The fidelity $\tilde{F}(\infty)$ is degraded by the width of the distribution $\varepsilon$, but the dependence is weaker than linear.

We have to be careful about the probability $\tilde{P}(N)$ for successful extraction of the state $\tilde{\rho}(N)$.
It asymptotically decays as 
\begin{equation}
\tilde{P}(N)\sim\tilde{\lambda}_0^N\Tr_{AB}\{\tilde{\Pi}_0\rho_0\}
\quad\mathrm{for\ large}\ N,
\label{eqn:DecayPF}
\end{equation}
where $\tilde{\Pi}_0$ is the eigenprojection of $\tilde{\mathcal{T}}$ belonging to its largest eigenvalue $\tilde{\lambda}_0$.
In the ideal case (with $\varepsilon=0$), the largest eigenvalue is $\lambda_0=1$ (Fig.\ \ref{fig:Eigenvalues}) and the probability ceases to decay as (\ref{eqn:Prob}).
This is not the case for a finite width $\varepsilon>0$: the largest eigenvalue $\tilde{\lambda}_0$ deviates from unity, as shown in Fig.\ \ref{fig:Fluctuation}.

As the figures in Fig.\ \ref{fig:Fluctuation} show, a larger resonant wave vector $k$ and a smaller coupling constant $g$ would be preferred to reduce the effect of the band width $\varepsilon$.
However, as mentioned in Sec.\ \ref{sec:Fidelity}, the extraction is slow in such a regime, and the probability $\tilde{P}(N)$ would decay out completely as  (\ref{eqn:DecayPF}), before a state with a high fidelity is extracted.
In Fig.\ \ref{fig:FPF}, the fidelity $\tilde{F}(N)$ and the probability $\tilde{P}(N)$ are plotted as functions of the number of transmissions $N$, for a finite $\varepsilon$.
The probability $\tilde{P}(N)$ keeps on decaying, while the fidelity $\tilde{F}(N)$ goes up quickly and a good fidelity would be expected before the probability completely decays to zero.

\section{Summary}
We have presented a scheme for the extraction of entanglement in two noninteracting fixed qubits, via repetition of resonant transmission of ancilla (mediator) qubit through the target.
The resonant transmission works as a filter and the singlet state is extracted in the target qubits from an arbitrary given state, without initial preparation of the target.
Neither the preparation nor the post-selection of the spin state of the ancilla is required.

Concise and explicit expressions of the transmission and reflection coefficients, $T$ and $R$, are presented, and concrete analytical proofs have been given to the convergence to the target entangled state.
The effect of the finite size of the incident wave packet or the fluctuation of the incident momentum of the ancilla has also been investigated.

Interestingly, we can control the spin state without manipulating the spin degree of freedom.
The fact that it does not resort to a spin-resolved detection would be a nice feature from a practical point of view.

\ack
This work is supported by a Special Coordination Fund for Promoting Science and Technology and the Grant-in-Aid for Young Scientists (B) (No.\ 21740294) both from the Ministry of Education, Culture, Sports, Science and Technology, Japan,
by the bilateral Italian-Japanese Projects II04C1AF4E on ``Quantum Information, Computation and Communication'' of the Italian Ministry of Education, University and Research, and
by the Joint Italian-Japanese Laboratory on ``Quantum Information and Computation'' of the Italian Ministry for Foreign Affairs.

\bigskip\noindent
\textit{Note added.} After the submission of the manuscript, a related work \cite{ref:mmmm} appeared.

\appendix
\section{Scattering Matrix}
We sketch the calculation of the scattering matrix elements (\ref{eqn:SmatrixElements}) and the derivation of the operators $T$ and $R$ in (\ref{eqn:TRa}) and (\ref{eqn:TRb}).

Let $\ket{k\zeta}$ denote an eigenstate of the free Hamiltonian $H_0=p^2/2m$,
\begin{equation}
H_0\ket{k\zeta}=E_k\ket{k\zeta},\qquad
E_k=\frac{\hbar^2k^2}{2m},
\end{equation}
and $\ket{\Psi_k\zeta}$ an eigenstate of the total Hamiltonian $H=H_0+V$ belonging to the same eigenvalue $E_k$,
\begin{equation}
H\ket{\Psi_k\zeta}=E_k\ket{\Psi_k\zeta},
\end{equation}
where $\zeta$ represents the spin state of the three qubits $XAB$\@.
The scattering matrix $S$ is then given by \cite{ref:qpfesc-long}
\begin{equation}
\bra{k'\zeta'}S\ket{k\zeta}
=\delta(k-k')\delta_{\zeta\zeta'}
-2\pi i\delta(E_k-E_{k'})\bra{k'\zeta'}V\ket{\Psi_k\zeta}.
\label{eqn:SmatrixEle}
\end{equation}
The scattering state $\ket{\Psi_k\zeta}$ is formally the solution to the Lippmann-Schwinger equation
\begin{equation}
\ket{\Psi_k\zeta}
=\ket{k\zeta}+\frac{1}{E_k-H_0+i0^+}V\ket{\Psi_k\zeta}.
\end{equation}
In the coordinate representation, it reads
\begin{equation}
\bracket{x}{\Psi_k\zeta}
=\bracket{x}{k\zeta}-\int dx'\,G_k(x-x')\frac{2m}{\hbar^2}V(x')\bracket{x'}{\Psi_k\zeta},
\label{eqn:LippmannSchwinger}
\end{equation}
with Green's function
\begin{equation}
G_k(x)=\frac{\hbar^2}{2m}\int \frac{dq}{2\pi}\frac{e^{iqx}}{E_q-E_k-i0^+}
=\frac{i}{2|k|}e^{i|kx|}.
\end{equation}
We assume $k>0$ henceforth (the left-incident problem).
For the present problem, the potential is
\begin{equation}
V(x)=g(\bm{\sigma}^{(X)}\cdot\bm{\sigma}^{(A)})\delta(x+d/2)
+g(\bm{\sigma}^{(X)}\cdot\bm{\sigma}^{(B)})\delta(x-d/2),
\end{equation}
and the Lippmann-Schwinger equation (\ref{eqn:LippmannSchwinger}) yields \cite{ref:qpfesc-long}
\begin{eqnarray}
&{-i}\Omega(\bm{\sigma}^{(X)}\cdot\bm{\sigma}^{(A)})
\bracket{-d/2}{\Psi_k\zeta}
\nonumber\\
&\qquad\ %
=\frac{1}{1-R_AR_Be^{2ikd}}R_A
[
\bracket{-d/2}{k\zeta}
+e^{ikd}R_B
\bracket{d/2}{k\zeta}
],
\label{eqn:SourceABa}
\end{eqnarray}
\begin{eqnarray}
&{-i}\Omega(\bm{\sigma}^{(X)}\cdot\bm{\sigma}^{(B)})
\bracket{d/2}{\Psi_k\zeta}
\nonumber\\
&\qquad\ %
=\frac{1}{1-R_BR_Ae^{2ikd}}R_B
[
\bracket{d/2}{k\zeta}
+e^{ikd}R_A
\bracket{-d/2}{k\zeta}
],
\label{eqn:SourceABb}
\end{eqnarray}
where
\begin{equation}
R_{A/B}=-\frac{i\Omega(\bm{\sigma}^{(X)}\cdot\bm{\sigma}^{(A/B)})}{1+i\Omega(\bm{\sigma}^{(X)}\cdot\bm{\sigma}^{(A/B)})},
\end{equation}
and $\Omega$ is defined in (\ref{eqn:AlphaBetaOmega}).
By inserting (\ref{eqn:SourceABa}) and (\ref{eqn:SourceABb}) into (\ref{eqn:SmatrixEle}), one gets the scattering matrix elements (\ref{eqn:SmatrixElements}) with
\begin{eqnarray}
T=T_Be^{ikd}\frac{1}{1-R_Ae^{ikd}R_Be^{ikd}}T_A,
\label{eqn:TRformal-a}
\\
R=R_A+T_Ae^{ikd}R_Be^{ikd}\frac{1}{1-R_Ae^{ikd}R_Be^{ikd}}T_A,
\label{eqn:TRformal-b}
\end{eqnarray}
where
\begin{equation}
T_{A/B}=1+R_{A/B}.
\end{equation}
Note that $T_{A/B}$ and $R_{A/B}$ describe the changes induced in the spin state of $XAB$ when $X$ is transmitted and reflected, respectively, by the single scatterer $A/B$.
It is clear from the power series expansions of (\ref{eqn:TRformal-a}) and (\ref{eqn:TRformal-b}),
\begin{eqnarray}
T=\sum_{n=0}^\infty T_Be^{ikd}(R_Ae^{ikd}R_Be^{ikd})^nT_A,
\\
R=R_A+\sum_{n=0}^\infty T_Ae^{ikd}R_Be^{ikd}(R_Ae^{ikd}R_Be^{ikd})^nT_A,
\end{eqnarray}
that $T$ and $R$ are the superpositions of all possible bouncing processes between $A$ and $B$.
By evaluating the inverse operators in (\ref{eqn:TRformal-a}) and (\ref{eqn:TRformal-b}) \cite{ref:qpfesc-long}, one obtains (\ref{eqn:TRa}) and (\ref{eqn:TRb}).

The following relations are useful for dealing with the operators appearing in (\ref{eqn:TRa}) and (\ref{eqn:TRb}):
\begin{eqnarray}
&P_\pm(\bm{\sigma}^{(A)}-\bm{\sigma}^{(B)})
=\pm P_\pm i(\bm{\sigma}^{(A)}\times\bm{\sigma}^{(B)})
\nonumber
\\
&\quad
=(\bm{\sigma}^{(A)}-\bm{\sigma}^{(B)})P_\mp
=\pm i(\bm{\sigma}^{(A)}\times\bm{\sigma}^{(B)})P_\mp
\nonumber
\\
&\quad
=\frac{1}{2}[
(\bm{\sigma}^{(A)}-\bm{\sigma}^{(B)})
\pm i(\bm{\sigma}^{(A)}\times\bm{\sigma}^{(B)})
].
\end{eqnarray}

\end{document}